\documentclass[twocolumn,secnumarabic,amssymb, nobibnotes, aps, pra,showpacs,preprintnumbers]{revtex4-1}
\usepackage{amssymb}
\usepackage{array}
\usepackage{graphicx}
\usepackage{amsmath}
\usepackage{amsthm}
\usepackage{amsopn}
\def\uno{\mbox{1 \kern-.59em {\rm l}}}

\def\beq{\begin{equation}}
\def\eeq{\end{equation}}
\def\bea{\begin{eqnarray}}
\def\eea{\end{eqnarray}}

\begin{document}

\title{Phase transition and thermodynamic stability in extended phase space and charged Ho\v{r}ava-Lifshitz black holes}

\author{Mohammad Bagher Jahani Poshteh$^{1}$}
\email{mb.jahani@iasbs.ac.ir}
\author{Nematollah Riazi$^{2}$}
\email{n\underline{\space}riazi@sbu.ac.ir}
\affiliation{
$^1$Institute for Advanced Studies in Basic Sciences (IASBS), Zanjan 45195-1159, Iran\\
$^2$Department of Physics, Shahid Beheshti University, Tehran 19839, Iran}

\begin{abstract}
For charged black holes in Ho\v{r}ava-Lifshitz gravity, a second order phase transition takes place in extended phase space where the cosmological constant is taken as thermodynamic pressure. We relate the second order nature of phase transition to the fact that the phase transition occurs at a sharp temperature and not over a temperature interval. Once we know the continuity of the first derivatives of the Gibbs free energy, we show that all the Ehrenfest equations are readily satisfied. We study the effect of the perturbation of the cosmological constant as well as the perturbation of the electric charge on thermodynamic stability of Ho\v{r}ava-Lifshitz black hole. We also use thermodynamic geometry to study phase transition in extended phase space. We investigate the behavior of scalar curvature of Weinhold, Ruppeiner, and Quevedo metric in extended phase space of charged Ho\v{r}ava-Lifshitz black holes. It is checked if these curvatures could reproduce the result of specific heat for the phase transition.
\end{abstract}


\maketitle

\newpage
\section{Introduction}
%%%%%%%%%%%%%%%%%%%%%%%%%%%%%%%%%%%%%%%%%%%%%%%%%%%%%

Ever since the pioneering works of Hawking \cite{bhexp} and Bekenstein \cite{bhent}, black hole thermodynamics has been a fascinating field of research in gravitational physics. The discovery of phase transition between Schwarzschild-AdS and thermal AdS \cite{hptr} and its corresponding phase transition in the boundary conformal field theory \cite{adscft} have attracted much attention to the subject of phase transition in black hole thermodynamics \cite{lis11,lis12,lis13,lis14,lis15,lis16,lis17,lis18,lis19,lis110}.

Recently, there has been interest in studying black hole thermodynamics in extended phase space. By taking cosmological constant as thermodynamic pressure \cite{traschen,dolan5023} and its conjugate quantity as thermodynamic volume, the first law of black hole thermodynamics would be complete. Investigations on black holes in extended phase space have shown that some black holes possess a critical behavior in analogy with the $P-V$ criticality of van der Waals fluid \cite{kubizpv,mann6251,mannsherkat1}. The authors of \cite{moliuprd,moepl,moliuplb} have checked the validity of original Ehrenfest equations at the critical point to find the nature of phase transition. For original Ehrenfest equations to be useful for charged (rotating) black holes, these authors take the charge (angular momentum) of the black hole to be constant. In this way, the work term $\Phi dQ$ $(\Omega dJ)$ is omitted from the first law and one could find the same Ehrenfest equations as in ordinary thermodynamics.

In this paper, we investigate charged Ho\v{r}ava-Lifshitz black holes in extended phase space. Ho\v{r}ava-Lifshitz gravity is a UV complete non-relativistic power-counting renormalizable theory of gravity proposed by Ho\v{r}ava \cite{horava1,horava2}, which reduces to Einstein's general relativity in the IR limit. Its black hole solutions were presented in \cite{lmp,ccoprd,ks,ghodsi} and the thermodynamics of these black holes has been investigated in \cite{myung,castillo,majhiplb,mojhep,suresh,cao5044,majhicqg}. The authors of \cite{cao5044} study the thermodynamics and phase transition of charged Ho\v{r}ava-Lifshitz black holes in non-extended phase space. They examine different topologies. The most interesting case is that of hyperbolic 2-space in which a phase transition takes place at the point where specific heat diverges. For uncharged Ho\v{r}ava-Lifshitz black hole, the thermodynamic quantities have been obtained in extended phase space \cite{sadeghi}. There, it is shown that the phase transition takes place only for hyperbolic space.

Our investigation on the extended charged Ho\v{r}ava-Lifshitz black holes is performed in three different ensembles; canonical ensemble in which both charge and pressure (cosmological constant) are taken to be constant; grand canonical ensemble with respect to electric charge, in which the electric charge exchanges between the black hole and its environment but pressure is fixed; and an ensemble in which both the charge and pressure are allowed to change. The specific heat has a divergent point in all of these ensembles which indicates a phase transition. We use the generalized version of Ehrenfest equations, developed recently in \cite{mine}, to further study the point at which the specific heat diverges. We show that, given the continuity of the first derivatives of the Gibbs free energy, all of the extended Ehrenfest equations are readily satisfied at the divergent point of the specific heat.

Our study of charged Ho\v{r}ava-Lifshitz black hole, also shows that the phase transition occurs at a sharp temperature at which the specific heat diverges. We prove that this is a sufficient condition for the Prigogine-Defay (PD) ratio to be equal to unity. So, we prove that the divergence of the specific heat at a sharp temperature is adequate for classifying the phase transition as second order. We also investigate local thermodynamic stability of charged Ho\v{r}ava-Lifshitz black hole under perturbations of cosmological constant (pressure) and charge of the black hole.

We use the geometric approach to investigate phase transition in extended phase space. Weinhold \cite{weinhold} and Ruppeiner \cite{rupp79} have introduced a Riemannian structure for thermodynamic equilibrium state space, but their proposed metrics are not invariant under Legendre transformation \cite{que06}. By invoking Legendre invariance, Quevedo presented the formalism of geometrothermodynamics \cite{que06}, which permits the construction of Legendre invariant metric. We investigate the scalar curvature of Weinhold, Ruppeiner, and Quevedo metric in the ensemble with varying charge and pressure to see if they produce the same result as the specific heat.

The outline of this paper is as follows. In Sect. \ref{sec2}, the thermodynamics and phase transition of charged extended Ho\v{r}ava-Lifshitz black hole is investigated. In Sect. \ref{sec3}, we prove that all of the extended Ehrenfest equations are valid at the divergent point of the specific heat. In Sect. \ref{sec4} we obtain conditions under which the Ho\v{r}ava-Lifshitz black hole is stable under the perturbations of the pressure and the charge. The study of the scalar curvature in the geometry of the extended thermodynamic phase space is presented in Sect. \ref{sec5}. We give our conclusions in Sect. \ref{seccon}. The derivation of the extended Ehrenfest equations is given in an Appendix.

%%%%%%%%%%%%%%%%%%%%%%%%%%%%%%%%%%%%%%%%%%%%%%%%%%%%%%%%%%%%%%%%%%%%%%%%%%%%%%%%%%%%%%%%%%%%%%%%%%%%%%%%%%%%%%%%%%%%%%%%%

\section{Thermodynamic phase transition of charged Ho\v{r}ava-Lifshitz black hole}
\label{sec2}

Ho\v{r}ava-Lifshitz theory is a power-counting renormalizable theory of gravity with an anisotropy between space and time \cite{horava1,horava2}. It can be treated as a candidate for a quantum gravity theory. In the last few years, its black hole solutions and the corresponding thermodynamics have been studied in literature \cite{lmp,ccoprd,ks,ghodsi,myung,castillo,majhiplb,mojhep,suresh,cao5044,majhicqg}.

In ADM formalism the line element of spacetime is written as
\begin{equation}
ds^{2}=-N^{2}dt^{2}+g_{ij}(dx^{i}+N^{i}dt)(dx^{j}+N^{j}dt),
\end{equation}
where $N$ and $N^{i}$ are the lapse and shift functions, respectively. Also $g_{ij}$ denotes the metric of fixed-time leaves of spacetime foliation with extrinsic curvature defined by
\begin{equation}
K_{ij}=\frac{1}{2N}(\dot{g}_{ij}-\nabla_{i}N_{j}-\nabla_{j}N_{i}),
\end{equation}
here the dot denotes the derivative with respect to time and covariant derivatives are with respect to $g_{ij}$. The Lagrangian of Ho\v{r}ava-Lifshitz gravity is
\begin{eqnarray}
{\mathcal L}&=&{\mathcal L}_{0}+{\mathcal L}_{1}+{\mathcal L}_{em}, \label{hllag} \\
{\mathcal L}_{0}&=&\sqrt{g}N\{\frac{2}{\kappa^{2}}(K_{ij}K^{ij}-\lambda K^{2})+\frac{\kappa^{2}\mu^{2}(\Lambda R-3\Lambda^{2})}{8(1-3\lambda)}\}, \nonumber \\
{\mathcal L}_{1}&=&\sqrt{g}N\{\frac{\kappa^{2}\mu^{2}(1-4\lambda)}{32(1-3\lambda)}R^{2}-\frac{\kappa^{2}}{2\omega^{4}}(C_{ij}-\frac{\mu\omega^{2}}{2}R_{ij})  \nonumber \\ &&\times(C^{ij}-\frac{\mu\omega^{2}}{2}R^{ij})\}, \nonumber \\
{\mathcal L}_{em}&=&p^{i}\dot{A}_{i}-\frac{1}{2}N(\frac{\alpha}{\sqrt{g}}p^{i}p_{i}+\frac{\sqrt{g}}{2\alpha}F_{ij}F^{ij})+\phi p^{i}_{,i}, \nonumber
\end{eqnarray}
where $\lambda$, $\kappa$, $\mu$ and $\omega$ are coupling constants. $g$ is the determinant of the spatial metric $g_{ij}$ and $C_{ij}$ is the Cotton tensor defined by
\begin{equation}
C^{ij}=\varepsilon^{ikl}\nabla_{k}(R^{i}_{l}-\frac{1}{4}R\delta^{i}_{l}).
\end{equation}
$\Lambda$ is the cosmological constant which we take to be negative. Also {\bf p} is the momentum conjugate to the spatial components of the Maxwell field $(\phi, {\bf A})$ and $\alpha=-16/(\kappa^{2}\mu^{2}\Lambda)$.

Topological black holes are of the form
\begin{equation}
ds^{2}=-\tilde{N}(r)^{2}f(r)dt^{2}+\frac{dr^{2}}{f(r)}+r^{2}d\Omega_{k}^{2}, \label{tople}
\end{equation}
where $d\Omega_{k}^{2}$ is the line element of two-dimensional Einstein space with constant scalar curvature $2k$. By plugging in the metric (\ref{tople}) into the Lagrangian (\ref{hllag}), and using the new variable $x=\sqrt{-\Lambda}r$, we could write out the action as \cite{ccoprd}
\begin{equation}
I=\frac{\kappa^{2}\mu^{2}\sqrt{-\Lambda}\Omega_{k}}{16}\int dt dx (\tilde{N}(U^{'}-\frac{1}{2}x^{2}\tilde{p}^{2})+\phi(x^{2}\tilde{p})^{'})+B, \label{action}
\end{equation}
with $\tilde{p}=\alpha p^{r}/(\sqrt{-\Lambda \gamma}r^{2})$ where $\gamma$ is the determinant of the 2-dimensional Einstein space with volume $\Omega_{k}$. $B$ is a boundary term and
\begin{equation}
U=x^{3}-2x(f-k)+\frac{(f-k)^{2}}{x}. \label{u}
\end{equation}
By varying the action (\ref{action}), one obtains the equations of motion
\begin{equation}
\tilde{N}^{'}=0, \qquad (x^{2}\tilde{p})^{'}=0, \qquad U^{'}=\frac{1}{2}x^{2}\tilde{p}^{2}, \label{eom}
\end{equation}
which result in
\begin{eqnarray}
f(r)&=&k+x^{2}-\sqrt{c_{0}x-\frac{q^{2}}{2}}, \\
c_{0}&=&\frac{2k^{2}+q^{2}+4kx_{+}^{2}+2x_{+}^{4}}{2x_{+}}. \nonumber
\end{eqnarray}
$x_{+}$ is the value of $x$ for which $f(r)=0$. $q$ is related to the charge of the black hole as it will be seen shortly. Also, $\tilde{N}(r)$ is constant and it could be set to one by rescaling the time coordinate. Mass, charge, temperature, and entropy of the black hole are respectively \cite{ccoprd}
\begin{eqnarray}
M&=&\frac{\kappa^{2}\mu^{2}\Omega_{k}\sqrt{-\Lambda}}{16}c_{0}, \label{nxm}\\
Q&=&\frac{\kappa^{2}\mu^{2}\Omega_{k}\sqrt{-\Lambda}}{16}q, \label{nxq}\\
T&=&\frac{\sqrt{-\Lambda}(3x_{+}^{4}+2kx_{+}^{2}-k^{2}-q^{2}/2)}{8\pi x_{+}(k+x_{+}^{2})}, \label{nxt}\\
S&=&\frac{\pi \kappa^{2}\mu^{2}\Omega_{k}}{4}(x_{+}^{2}+2k\ln x_{+})+S_{0} \label{nxs},
\end{eqnarray}
in which $S_{0}$ is an integration constant.

In a recent study \cite{nopv}, phase space of the charged Ho\v{r}ava-Lifshitz black hole has been extended to include the thermodynamic pressure $P=-\Lambda/8\pi$ and its conjugate quantity, referred to as thermodynamic volume $V$. So we may eliminate $\Lambda$ in favor of $P$ and write Eqs. (\ref{nxm}), (\ref{nxt}), and (\ref{nxs}) as
\begin{eqnarray}
M&=&\frac{\pi P(k+8\pi Pr_{+}^2)^{2}+Q^{2}}{4\pi Pr_{+}}, \label{m}\\
T&=&\frac{\pi P(k+8\pi Pr_{+}^{2})(-k+24\pi Pr_{+}^{2})-Q^{2}}{8\pi^{2}Pr_{+}(k+8\pi Pr_{+}^{2})}, \label{t}\\
S&=&\pi(8\pi Pr_{+}^{2}+2k\ln(\sqrt{8\pi P}r_{+}))+S_{0}, \label{s}
\end{eqnarray}
where we have set $\kappa^{2}\mu^{2}\Omega_{k}=4$ for simplicity.

By introducing thermodynamic pressure, mass of the black hole would be associated with the enthalpy and the first law of black hole thermodynamics takes the form \cite{traschen}
\begin{equation}
dM=TdS+\Phi dQ+VdP. \label{1stlaw}
\end{equation}
Using this equation and Eqs. (\ref{m}) and (\ref{s}), thermodynamic volume would be obtained as
\begin{eqnarray}
V&=&(\frac{\partial M}{\partial P})_{S,Q}=(\frac{\partial M}{\partial P})_{r_{+},Q}+(\frac{\partial M}{\partial r_{+}})_{P,Q}(\frac{\partial r_{+}}{\partial P})_{S,Q}  \nonumber \\
&=&\frac{\pi P(k+8\pi Pr_{+}^{2})^{2}-Q^{2}}{8\pi P^{2}r_{+}}, \label{v}
\end{eqnarray}
where in the last equality we have used the identity
\begin{equation}
(\frac{\partial X}{\partial Y})_{Z}=-(\frac{\partial Z}{\partial Y})_{X}(\frac{\partial X}{\partial Z})_{Y}. \label{iden}
\end{equation}
Also from Eqs. (\ref{1stlaw}) and (\ref{m}) we can find the electric potential
\begin{equation}
\Phi=(\frac{\partial M}{\partial Q})_{S,P}=(\frac{\partial M}{\partial Q})_{r_{+},P}=\frac{Q}{2\pi Pr_{+}}, \label{phi}
\end{equation}
in which we have used the fact that the entropy (\ref{s}) is just a function of $r_{+}$ and $P$ and not the charge $Q$.

Here we investigate the possibility of phase transition through studying the behavior of the specific heat in three different ensembles. By taking the pressure to be constant, one would have canonical and grand canonical ensembles with respect to the elctric charge. In these cases our analysis of the specific heat reduces to that of \cite{cao5044} in non-extended phase space. We briefly give the results. First consider the ensemble in which both pressure and charge are constant. By using Eqs. (\ref{t}) and (\ref{s}), we obtain the specific heat at constant pressure and charge
\begin{widetext}
\begin{align}
C_{P,Q}=T(\frac{\partial S}{\partial T})_{P,Q}=\frac{2\pi(k+8\pi Pr_{+}^{2})^{2}(\pi P(k+8\pi Pr_{+}^{2})(-k+24\pi Pr_{+}^{2})-Q^{2})}{(k+24\pi Pr_{+}^{2})(\pi P(k+8\pi Pr_{+}^{2})^{2}+Q^{2})}.
\label{c}
\end{align}
\end{widetext}

It is obvious from Eq. (\ref{c}) that for the case of hyperbolic 2-space with $k=-1$, the specific heat diverges at $r_{+,c}=1/\sqrt{24\pi P}$, which is independent of the charge. In the next section, we prove that the continuity of the entropy, volume, and electric potential, along with the divergence of the specific heat, is sufficient for classifying the phase transformation as second order. By putting $r_{+,c}$ for event horizon radius in (\ref{t}) with $k=-1$, the temperature at which the phase transition takes place is found to be $T_{c}=(3/\sqrt{6\pi^{3}P})(\frac{3}{4}Q^{2}+\pi P)$.

\begin{figure}[t]
	\centering
	\includegraphics[scale=.9]{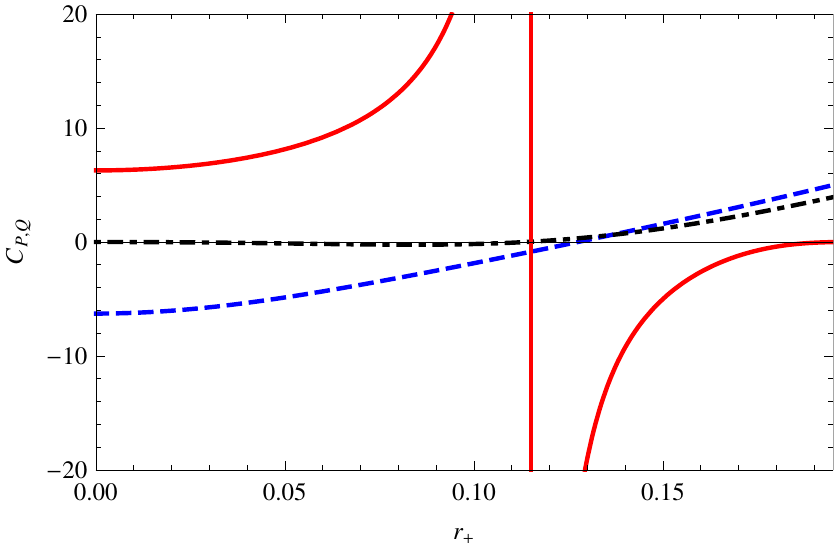}
	\caption[]{Specific heat $C_{P,Q}$ as a function of horizon radius $r_{+}$ for spherical [blue (dashed) line], flat [black (dashed-dotted) line], and hyperbolic [red (solid) line] 2-spaces with $Q=1$ and $P=1$.}
	\label{cplot}
\end{figure}

In Fig. \ref{cplot}, the specific heat $C_{P,Q}$ is depicted as a function of horizon radius $r_{+}$ for three cases of spherical ($k=1$), flat ($k=0$), and hyperbolic ($k=-1$) 2-spaces. There is no divergence point for $k=1$ or $0$. This is different from the behavior of charged black holes in general relativity in which the specific heat only diverges for black holes with spherical horizon ($k=1$) \cite{kubizpv,brill}. One could further investigate the relations between topological black holes of Ho\v{r}ava-Lifshitz theory and that of Einstein gravity.

It was first pointed out in \cite{ccoprd} that there exist a duality between temperature of topological black holes in Ho\v{r}ava-Lifshitz gravity with $k=-1, 0$, and $1$ respectively to that of topological black holes in general relativity with  $k=1, 0$, and $-1$. This fact could be traced back to the presence of higher spatial derivative in the Lagrangian (\ref{hllag}) of Ho\v{r}ava-Lifshitz theory. Without the higher derivative terms, the last term in the right hand side of Eq. (\ref{u}) will be omitted. By solving the equations of motion (\ref{eom}) with $U=x^{3}-2x(f-k)$ one obtains
\begin{eqnarray}
f(r)&=&k-\frac{c}{2x}+\frac{q^{2}}{4x^{2}}+\frac{x^{2}}{2}, \label{fgr} \\
c&=&\frac{q^{2}+4kx_{+}^{2}+2x_{+}^{4}}{2x_{+}}, \nonumber
\end{eqnarray}
which is of the same form as the metric of Reissner-Nordstr\"{o}m-AdS black hole \cite{brill}. By using Eq. (\ref{fgr}) one could obtain the specific heat of the black hole in the \textit{low-derivative} theory. The duality which is pointed in \cite{ccoprd} would then be obvious.

If we let the charge exchange between the black hole and its environment (and take the pressure to be constant) we would have grand canonical ensemble with respect to the electric charge. By substituting $Q$ from Eq. (\ref{phi}) into Eq. (\ref{t}), we find
\begin{equation}
T=\frac{Pr_{+}}{2}(6-\frac{\Phi^{2}}{k+8\pi Pr_{+}})-\frac{k}{8\pi r_{+}}.
\end{equation}
By using this equation and Eq. (\ref{s}), we obtain the specific heat at constant pressure and electric potential
\begin{widetext}
\begin{align}
C_{P,\Phi}=T(\frac{\partial S}{\partial T})_{P,\Phi}=\frac{2\pi(8\pi Pr_{+}^{2}+k)^{2}((24\pi Pr_{+}^{2}-k)(8\pi Pr_{+}^{2}+k)-4\pi P\Phi^{2}r_{+}^{2})}{k^{3}+4k\pi Pr_{+}^{2}(10k-\Phi^{2})+32\pi^{2}P^{2}r_{+}^{4}(14k+\Phi^{2}+48\pi Pr_{+}^{2})}.
\label{cgc}
\end{align}
\end{widetext}

\begin{figure}[t]
	\centering
	\includegraphics[scale=.9]{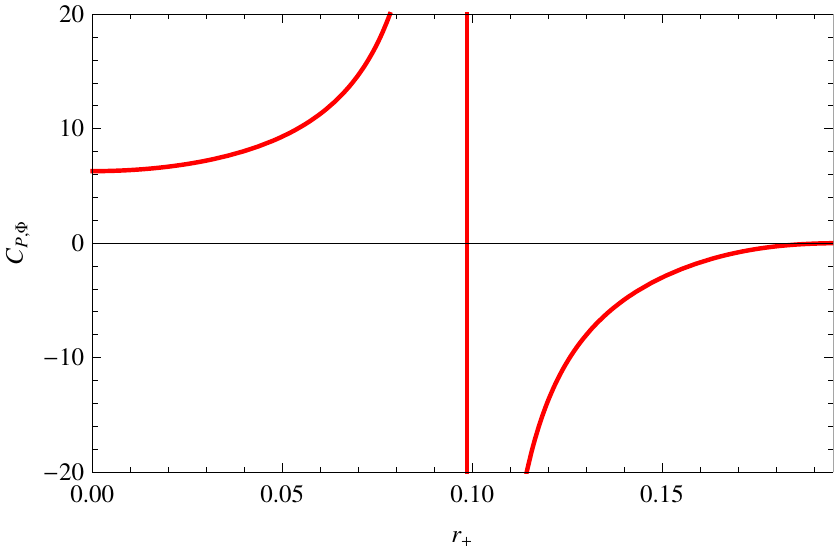}
	\caption[]{Specific heat $C_{P,\Phi}$ as a function of horizon radius $r_{+}$ for $k=-1$, $P=1$, and $\Phi=1$.}
	\label{cgcplot}
\end{figure}

Equation (\ref{cgc}) has a divergent point for $k=-1$ which shows a second order phase transition. It has also two divergent points for the case $k=1$, but none of them lies in the physical region, since the Hawking temperature is negative there. We plot the specific heat $C_{P,\Phi}$ in Fig. \ref{cgcplot}.

\begin{figure}[t]
	\centering
	\includegraphics[scale=.9]{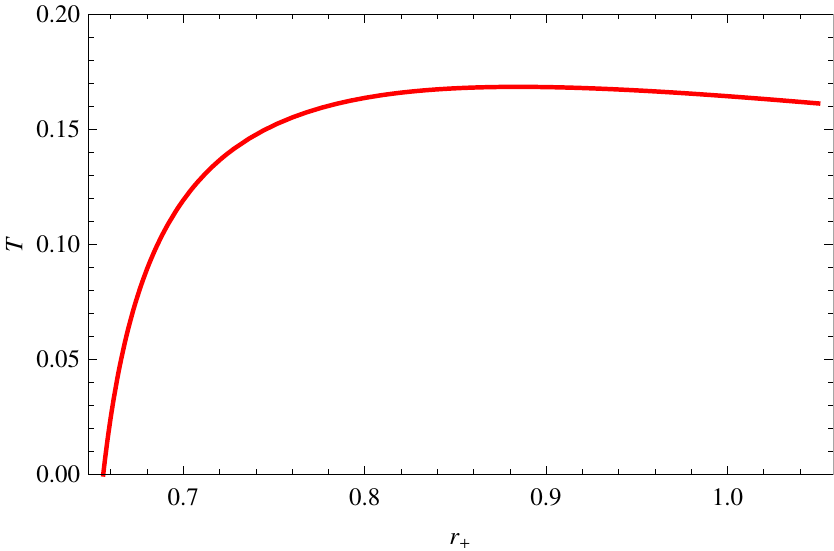}
	\caption[]{Temperature $T$ as a function of horizon radius $r_{+}$ for $k=-1$, $\Phi=1$, and $V=-1$.}
	\label{tvphip}
\end{figure}

One can as well study the thermodynamics of charged extended black holes in an ensemble that both charge and pressure could vary. By using Eqs. (\ref{v}) and (\ref{phi}) we can write temperature and entropy as a function of horizon radius $r_{+}$, electric potential $\Phi$, and thermodynamic volume $V$. The explicit expression for temperature in terms of $r_{+}$, $\Phi$, and $V$ is
\begin{equation}
T=\frac{(\sigma+\Phi^{2}-4k)(6-\frac{4\Phi^{2}}{\sigma+\Phi^{2}})-8k}{64\pi r_{+}},
\label{tvphi}
\end{equation}
where
\begin{equation}
\sigma=\frac{1}{\pi r_{+}}(2V+\sqrt{(\pi r_{+}\Phi^{2}+2V)(2V+\pi r_{+}(\Phi^{2}-8k))}).
\end{equation}
We have plotted the temperature (\ref{tvphi}) in Fig. \ref{tvphip}.

\begin{figure}[t]
	\centering
	\includegraphics[scale=.9]{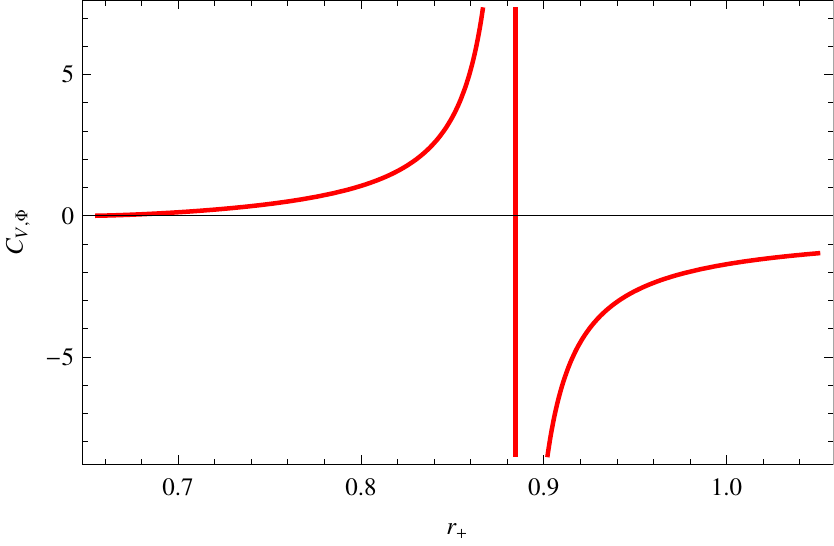}
	\caption[]{Specific heat $C_{\Phi,V}$ as a function of horizon radius $r_{+}$ for $k=-1$, $\Phi=1$, and $V=-1$.}
	\label{cvphi}
\end{figure}

Now we can calculate the specific heat $C_{\Phi,V}=T(\frac{\partial S}{\partial T})_{\Phi,V}$. The explicit equation for $C_{\Phi,V}$ is too lengthy to write here but we have plotted it for the case $k=-1$ in Fig. \ref{cvphi} with $\Phi=1$, and $V=-1$. At $r_{+}=0.65586$ the specific heat and the temperature are zero, indicating the extremal black hole. The specific heat diverges at $r_{+}=0.88443$ which is indicative of a second order phase transition. It should be noted that the specific heat $C_{\Phi,V}$ does not diverge for positive values of the thermodynamic volume. We do not have a clear explanation for the sign of the thermodynamic volume yet. But, as we will show in Sect. \ref{sec4}, local stability restricts the value of thermodynamic volume to be negative.

\section{Application of Ehrenfest scheme to black holes in extended phase space} \label{sec3}

\begin{table*}
	\begin{center}
		\caption{Extended Ehrenfest equations for charged black holes.}
		\begin{tabular}{@{}lllll}
			\hline\hline
			&&\\
			$(\frac{\partial P}{\partial T})_{S,Q}=\frac{\Delta C_{P,Q}}{TV\Delta\alpha}$ &
			$(\frac{\partial P}{\partial T})_{V,Q}=\frac{\Delta\alpha}{\Delta\kappa}$ &
			$(\frac{\partial P}{\partial T})_{\Phi,Q}=-\frac{\Delta\alpha^{'}}{\Delta\kappa^{'}}$\\
			&&\\
			$(\frac{\partial P}{\partial Q})_{T,S}=-\frac{\Phi\Delta\alpha^{'}}{V\Delta\alpha}$ &
			$(\frac{\partial P}{\partial Q})_{T,V}=\frac{\Phi\Delta\kappa^{'}}{V\Delta\kappa}$ &
			$(\frac{\partial P}{\partial Q})_{T,\Phi}=-\frac{\Delta\chi}{\Delta\kappa^{'}}$\\
			&&\\
			$(\frac{\partial Q}{\partial T})_{S,P}=\frac{\Delta C_{P,Q}}{T\Phi\Delta\alpha^{'}}$&
			$(\frac{\partial Q}{\partial T})_{V,P}=-\frac{V\Delta\alpha}{\Phi\Delta\kappa^{'}}$&
			$(\frac{\partial Q}{\partial T})_{\Phi,P}=-\frac{\Delta\alpha^{'}}{\Delta\chi}$\\
			&&\\
			\hline
			&&\\
			$C_{P,Q}=T(\frac{\partial S}{\partial T})_{P,Q}$ &
			$\kappa=-\frac{1}{V}(\frac{\partial V}{\partial P})_{T,Q}$ &
			$\alpha^{'}=-\frac{1}{\Phi}(\frac{\partial S}{\partial Q})_{P,T}=\frac{1}{\Phi}(\frac{\partial \Phi}{\partial T})_{P,Q}$\\
			&&\\
			$\alpha=-\frac{1}{V}(\frac{\partial S}{\partial P})_{T,Q}=\frac{1}{V}(\frac{\partial V}{\partial T})_{P,Q}$ &
			$\kappa^{'}=\frac{1}{\Phi}(\frac{\partial V}{\partial Q})_{T,P}=\frac{1}{\Phi}(\frac{\partial \Phi}{\partial P})_{T,Q}$ &
			$\chi=\frac{1}{\Phi}(\frac{\partial\Phi}{\partial Q})_{T,P}$\\
			&&\\
			\hline\hline
		\end{tabular} \label{eqstab}
	\end{center}
\end{table*}

For black holes with three independent parameters (i.e.\ temperature $T$, pressure $P$, and charge $Q$), classical Ehrenfest equations should be generalized. In the Appendix, by using the method of \cite{mine} we derive nine Ehrenfest equations for extended charged black holes in an ensemble with fixed electric charge and pressure. These equations are collected together in Table \ref{eqstab}. In grand canonical ensemble with respect to the electric charge, extended Ehrenfest equations can be read off from Table \ref{eqstab} by replacing $Q$ by $-\Phi$ and $\Phi$ by $Q$. Besides, in an ensemble in which both charge and pressure could vary, extended Ehrenfest equations would be found by replacing $Q$ by $-\Phi$, $\Phi$ by $Q$, $P$ by $-V$, and $V$ by $P$.

In the limit of uncharged black hole, only the first equation
\begin{equation}
(\frac{\partial P}{\partial T})_{S}=\frac{\Delta C_{P}}{TV\Delta\alpha},
\end{equation}
and the fourth one,
\begin{equation}
(\frac{\partial P}{\partial T})_{V}=\frac{\Delta\alpha}{\Delta\kappa},
\end{equation}
would remain, which are identical to the classical Ehrenfest equations. Here $\Delta$ indicates the difference of the parameter before and after the transition.

In this section, we show that, given the continuity of the first derivatives of the Gibbs free energy, all of the extended Ehrenfest equations, including the classical ones, are satisfied at the point of the divergence of the specific heat.

By using the definition of the specific heat and volume expansion coefficient $\alpha$ from Table \ref{eqstab}, we have
\begin{eqnarray}
C_{P,Q}&=&T(\frac{\partial S}{\partial T})_{P,Q}=-T(\frac{\partial P}{\partial T})_{S,Q}(\frac{\partial S}{\partial P})_{T,Q} \nonumber\\
&=&TV\alpha(\frac{\partial P}{\partial T})_{S,Q}, \label{1sttr}
\end{eqnarray}
in which we have used the identity (\ref{iden}). Equation (\ref{1sttr}) could be transformed into
\begin{equation}
\Delta C_{P,Q}=TV\Delta\alpha(\frac{\partial P}{\partial T})_{S,Q}, \label{del1}
\end{equation}
which proves the validity of the first extended Ehrenfest equation.

Now, consider the definition of volume expansion coefficient $\alpha$ and isothermal compressibility $\kappa$. We have
\begin{equation}
V\alpha=(\frac{\partial V}{\partial T})_{P,Q}=-(\frac{\partial P}{\partial T})_{V,Q}(\frac{\partial V}{\partial P})_{T,Q}=V\kappa(\frac{\partial P}{\partial T})_{V,Q},
\end{equation}
which results in
\begin{equation}
\Delta\alpha=\Delta\kappa(\frac{\partial P}{\partial T})_{V,Q}. \label{del2}
\end{equation}
This proves the validity of the fourth extended Ehrenfest equation. Following the same procedure as above and using the identity (\ref{iden}), one can prove that all of the other extended Ehrenfest equations hold at the point of transition.

The importance of the first and the fourth extended Ehrenfest equations is that, by using them, we can obtain the PD ratio defined as \cite{pd}
\begin{equation}
\Pi=\frac{\Delta C_{P,Q}\Delta\kappa}{TV(\Delta\alpha)^{2}}.
\end{equation}
From Eqs. (\ref{del1}) and (\ref{del2}) we obtain
\begin{equation}
\Pi=(\frac{\partial P}{\partial T})_{S,Q}\bigg|_{c}(\frac{\partial T}{\partial P})_{V,Q}\bigg|_{c}.
\end{equation}
By ``c'' we mean that the derivaives have to be taken at the critical point. We have
\begin{equation}
(\frac{\partial T}{\partial P})_{V,Q}\bigg|_{c}=(\frac{\partial T}{\partial P})_{S,Q}\bigg|_{c}+(\frac{\partial T}{\partial S})_{P,Q}\bigg|_{c}(\frac{\partial S}{\partial P})_{V,Q}\bigg|_{c}. \label{forpd}
\end{equation}
Since the critical point is at the minimum temperature, we conclude that the second term in Eq. (\ref{forpd}) vanishes and we would obtain
\begin{equation}
\Pi=(\frac{\partial P}{\partial T})_{S,Q}\bigg|_{c}(\frac{\partial T}{\partial P})_{S,Q}\bigg|_{c}=1,
\end{equation}
which is the value of PD ratio for the equilibrium phase transition of second order. So, we have shown that the PD ratio is equal to one at the point where the specific heat diverges. The phase transitions for which the PD ratio is more than unity, are classified as glassy phase transitions \cite{jackle}. Also, for such phase transformations, not all of the Ehrenfest equations are satisfied. In fact, the fourth extended Ehrenfest equation does not hold for the glassy phase transition \cite{jackle,nie}.

Extended Ehrenfest equations give us nine relations between the variation of pressure/charge and other thermodynamic quantities at the point of second order phase transition. Let us resume the case of charged Ho\v{r}ava-Lifshitz black hole with $k=-1$. The explicit expressions of the parameters which appear on the right hand side of extended Ehrenfest equations are given in Table \ref{seec}.

\begin{table}[h]
	\begin{center}
		\caption{Parameters of the extended Ehrenfest equations for charged Ho\v{r}ava-Lifshitz black hole.}
		\begin{tabular}{@{}lllll}
			\hline
			&&\\
			$\alpha$ & $\frac{8\pi^{2}Pr_{+}(1-8\pi Pr_{+}^{2})^{2}(\pi P(8\pi Pr_{+}^{2}-1)(1+24\pi Pr_{+}^{2})-Q^{2})}{(24\pi Pr_{+}^{2}-1)(\pi^{2}P^{2}(1-8\pi Pr_{+}^{2})^{4}-Q^{4})}$ \\
			&&\\
			$\alpha'$ & $-\frac{8\pi^{2}Pr_{+}(1-8\pi Pr_{+}^{2})^{2}}{(24\pi Pr_{+}^{2}-1)(\pi P(1-8\pi Pr_{+}^{2})^{2}+Q^{2})}$ \\
			&&\\
			$\kappa$ & $\frac{\zeta}{P(24\pi Pr_{+}^{2}-1)(\pi^{2}P^{2}(1-8\pi Pr_{+}^{2})^{4}-Q^{4})}$ \\
			&&\\
			$\kappa'$ & $\frac{\pi(1-8\pi Pr_{+}^{2})^{2}-8\pi Q^{2}r_{+}^{2}}{(24\pi Pr_{+}^{2}-1)(\pi P(1-8\pi Pr_{+}^{2})^{2}+Q^{2})}$ \\
			&&\\
			$\chi$ & $\frac{2Q^{2}(1-8\pi Pr_{+}^{2})+(24\pi Pr_{+}^{2}-1)(\pi P(1-8\pi Pr_{+}^{2})^{2}+Q^{2})}{Q(24\pi Pr_{+}^{2}-1)(\pi P(1-8\pi Pr_{+}^{2})^{2}+Q^{2})}$ \\
			&&\\
			\hline
		\end{tabular}
		\label{seec}
	\end{center}
\end{table}

The auxiliary function $\zeta$, that appears in the expression of $\kappa$ is given by
\begin{eqnarray}
\zeta=2\pi PQ^{2}+16\pi Pr_{+}^{2}(\pi^{2}P^{2}-2\pi PQ^{2}-2Q^{4}) \nonumber\\
+Q^{4}-\pi^{2}P^{2}+128\pi^{3}P^{3}r_{+}^{4}(Q^{2}+\pi P(16\pi Pr_{+}^{2}-3) \nonumber\\
\times(1+16\pi Pr_{+}^{2}(6\pi Pr_{+}^{2}-1))). \nonumber
\end{eqnarray}

The parameters of Table \ref{seec}, have the same factor $(24\pi Pr_{+}^{2}-1)$ in their denominators which also appear in the denominator of the specific heat (\ref{c}) for the case $k=-1$. This factor forces all of these parameters diverge at the point of second order phase transition. Since the factor $(24\pi Pr_{+}^{2}-1)$ cancels from the nominator and denominator of the right hand side of extended Ehrenfest equations, these equations remain meaningful at the point of transition.

\section{Local thermodynamic stability} \label{sec4}

Consider a charged black hole in a thermal bath at constant temperature and electric potential and let the charge and pressure fluctuate. For the black hole to be in local thermodynamic stability, the second law must suppress the fluctuations of mass, charge, and pressure, so that the black hole does not evolve out of its equilibrium state. So by using the first law $\Delta M=T\Delta S+\Phi\Delta Q+V\Delta P$, one has for the locally thermodynamically stable black hole \cite{monteiro}
\begin{equation}
\Delta S-\frac{\Delta M-\Phi\Delta Q-V\Delta P}{T}<0.
\end{equation}
Upon expanding the entropy around its equilibrium value, and using the first law, we obtain for small fluctuations
\begin{widetext}
\begin{equation}
		\Delta S-\frac{\Delta M-\Phi\Delta Q-V\Delta P}{T}=\frac{1}{2}\frac{\partial^{2} S}{\partial X^{a}\partial X^{b}}\Delta X^{a}\Delta X^{b}, \qquad X^{a}=\{M, Q, P\}.
\end{equation}
So the black hole would be stable if and only if the Hessian $-\frac{\partial^{2} S}{\partial X^{a}\partial X^{b}}$ is positive definite. An equivalent condition for local thermodynamic stability is to demand positive definity of the Hessian $\frac{\partial^{2} M}{\partial E^{a}\partial E^{b}}$, in which $E^{a}=\{S, Q, P\}$ \cite{monteiro}. For the Hessian matrix $\frac{\partial^{2} M}{\partial E^{a}\partial E^{b}}$ to be positive definite, its determinant as well as the the determinant of $\frac{\partial^{2} M}{\partial \tilde{E}^{a}\partial \tilde{E}^{b}}$ ($\tilde{E}^{a}=\{S, Q\}$) and $\frac{\partial^{2} M}{\partial S^{2}}$ must be positive \cite{gabriel}.

Now consider the charged Ho\v{r}ava-Lifshitz black hole. We have
\begin{equation} \label{weinmet}
\left| \begin{array}{ccc}
\frac{\partial^{2} M}{\partial S^{2}} & \frac{\partial^{2} M}{\partial S\partial Q} & \frac{\partial^{2} M}{\partial S\partial P} \\
\frac{\partial^{2} M}{\partial S\partial Q} & \frac{\partial^{2} M}{\partial Q^{2}} & \frac{\partial^{2} M}{\partial Q\partial P} \\
\frac{\partial^{2} M}{\partial S\partial P} & \frac{\partial^{2} M}{\partial Q\partial P} & \frac{\partial^{2} M}{\partial P^{2}} \end{array} \right|=-\frac{A(r_{+}, Q, P)}{256P^{5}r_{+}^{3}(k+8\pi Pr_{+}^{2})^{3}},
\end{equation}
in which
\begin{eqnarray}
A(r_{+}, Q, P)=k(k^{4}\pi^{2}P^{2}+Q^{4})+16k^{2}\pi^{2}P^{2}r_{+}^{2}(k^{2}\pi P-2Q^{2})+128k\pi^{3}P^{3}r_{+}^{4}(3k^{2}\pi P-4Q^{2}) \nonumber \\
+2048\pi^{4}P^{4}r_{+}^{6}(4k^{2}\pi P-Q^{2})+69632k\pi^{6}P^{6}r^{8}+196608\pi^{7}P^{7}r^{10}. \nonumber
\end{eqnarray}
And
\begin{equation}
		\label{det1}
		\left| \begin{array}{cc}
		\frac{\partial^{2} M}{\partial S^{2}} & \frac{\partial^{2} M}{\partial S\partial Q} \\
		\frac{\partial^{2} M}{\partial S\partial Q} & \frac{\partial^{2} M}{\partial Q^{2}} \\
		\end{array} \right|=\frac{k^{3}\pi P-kQ^{2}+8\pi P r_{+}^{2}(5k^{2}\pi P+Q^{2})+448k\pi^{3}P^{3}r_{+}^{4}+1536\pi^{4}P^{4} r_{+}^{6}}{32\pi^{4}P^{2}r_{+}^{2}(k+8\pi Pr_{+}^{2})^{3}}.
\end{equation}
\end{widetext}
Also
\begin{equation}
\label{mss}
\frac{\partial^{2} M}{\partial S^{2}}=\frac{(k+24\pi Pr_{+}^{2})(Q^{2}+\pi P(k+8\pi Pr_{+}^{2})^{2})}{16\pi^{3}Pr_{+}(k+8\pi Pr_{+}^{2})^{3}}.
\end{equation}

\begin{figure}[t]
	\centering
	\includegraphics[scale=.9]{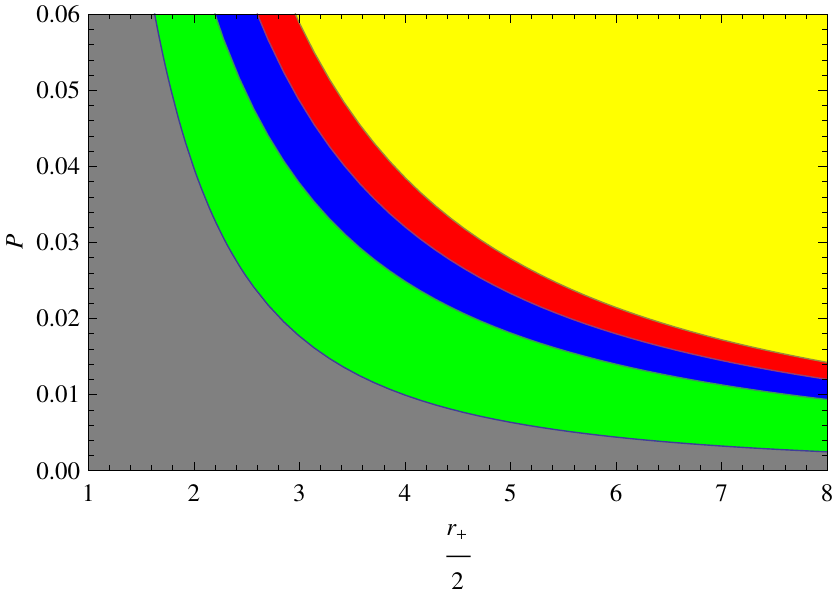}
	\caption[]{The region of phase space for which the charged Ho\v{r}ava-Lifshitz black hole is locally stable. We have considered hyperbolic $(k=-1)$ 2-spaces with $Q=1$. In the gray part (first region from left below) we have $T<0$, $V<0$. Also the black hole is unstable. In the green region (second from left below) the black hole is stable, but $T<0$ and $V<0$. In the blue region (middle one) the thermodynamic stability is satisfied, $T>0$ and $V<0$. In the red part (second region from right above) the black hole is unstable, $T>0$ and $V<0$. In the yellow region (the first one from right above) the black hole is unstable, $T>0$ and $V>0$.}
	\label{stab}
\end{figure}

The case of our interest is $k=-1$. So for the charged Ho\v{r}ava-Lifshitz black hole to be locally thermodynamically stable, the pressure must be limited from below and above by solutions of equations
\begin{equation}
8\pi Pr_{+}^{2}-1=0,
\end{equation}
and
\begin{eqnarray}
Q^{4}-\pi^{2}P^{2}(1-8\pi Pr_{+}^{2})^{2}(64\pi^{2}P^{2}r_{+}^{4}(48\pi Pr_{+}^{2}) \nonumber\\
-32Q^{2}r_{+}^{2}-1)=0.
\end{eqnarray}
For fluctuations of the charge around $Q=1$ the locally stable region of the black hole is shown in Fig. \ref{stab} for a specific range of parameters. The only physical region with positive temperature which is locally stable, is the blue part (the middle rigion) of the plot. For this region the thermodynamic volume is negative.

For an ensemble with fixed pressure and grand canonical with respect to electric charge the conditions for local stability reduces to positivity of Eqs. (\ref{det1}) and (\ref{mss}). This conditions is indeed equivalent to positivity of the specific heat (\ref{cgc}). This can be shown by defining the thermodynamic potential $\tilde{M}=M-\Phi Q$ and using the definition of the specific heat $C_{P,\Phi}^{-1}=\frac{1}{T}(\frac{\partial T}{\partial S})_{P,\Phi}=\frac{1}{T}(\frac{\partial^{2}\tilde{M}}{\partial S^{2}})_{P,\Phi}$. In an ensemble for which both pressure and electric charge are fixed, the stability condition reduces to positivity of Eq. (\ref{mss}) which is equivalent to positivity of the specific heat (\ref{c}).

\section{Singularities in thermodynamic geometry} \label{sec5}

\begin{figure}[t]
	\centering
	\includegraphics[scale=.9]{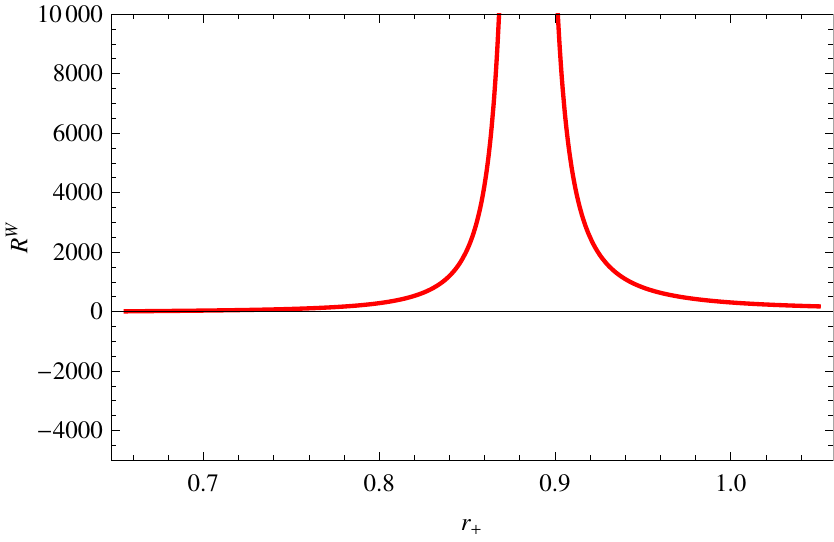}
	\caption[]{Scalar curvature of Weinhold metric as a function of horizon radius $r_{+}$ for $k=-1$, $\Phi=1$, and $V=-1$.}
	\label{wricci}
\end{figure}

The phase transition could also be investigated by incorporating geometric methods into thermodynamics. In an early work, Weinhold introduced Riemannian metric into equilibrium state space as the Hessian matrix of internal energy as a function of entropy and other extensive parameters \cite{weinhold}. In extended phase space it is convenient to work with enthalpy (mass) instead of internal energy. Since the enthalpy is related to the internal energy by the Legendre transformation $H=M=E+PV$, we define Weinhold metric as
\begin{equation}
g^{W}_{ab}=\frac{\partial^{2} M}{\partial N^{a}\partial N^{b}}, \qquad N^{a}=\{S,Q,P\}.
\end{equation}
$g^{W}_{ab}$ is indeed the metric (\ref{weinmet}). The scalar curvature of this metric is plotted in Fig. \ref{wricci}. Its explicit expression is too lengthy to write here. Weinhold curvature, $R^{W}$, diverges at $r_{+}=0.88443$ which is the divergent point of the specific heat. So $R^{W}$ could be used to find the point of the second order phase transition of Ho\v{r}ava-Lifshitz black holes in extended phase space.

\begin{figure}[t]
	\centering
	\includegraphics[scale=.9]{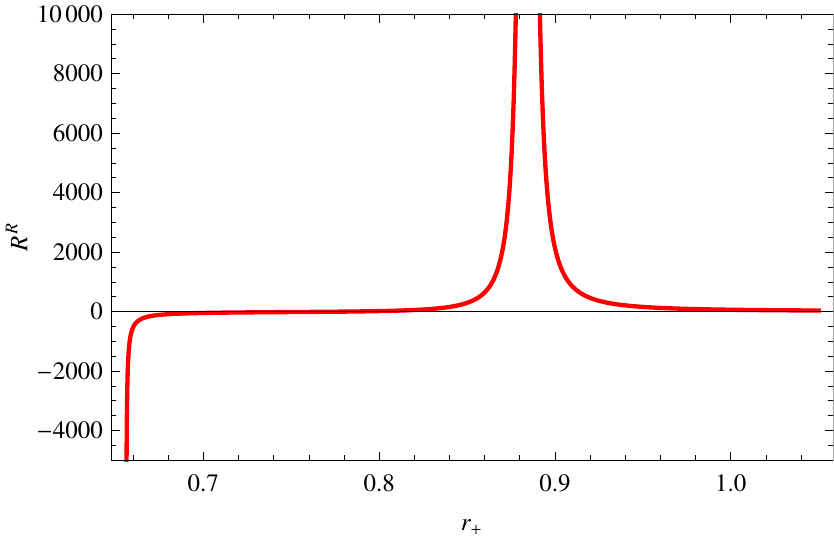}
	\caption[]{Scalar curvature of Ruppeiner metric as a function of horizon radius $r_{+}$ for $k=-1$, $\Phi=1$, and $V=-1$.}
	\label{rricci}
\end{figure}

A few years after Weinhold, Ruppeiner proposed a similar metric defined as the Hessian matrix of entropy, where the derivatives are taken with respect to internal energy and other extensive variables \cite{rupp79}. It can be shown \cite{salamon} that the line element of the Ruppeiner metric could be related to that of Weinhold via the conformal relation $ds_{R}^{2}=\frac{ds_{W}^{2}}{T}$, with the inverse temperature as the conformal factor. So by multiplying the conformal factor $\frac{1}{T}$ to the metric (\ref{weinmet}), we can calculate the scalar curvature of Ruppeiner metric, $R^{R}$, the plot of which is presented in Fig. \ref{rricci}. $R^{R}$ diverges at $r_{+}=0.88443$, at which the second order phase transition takes place. It also goes to infinity for the extremal black hole at $r_{+}=0.65586$.

An interesting feature of the Ruppeiner curvature is that it changes sign at $r_{+}=0.78348$ by passing through $R^{R}=0$. Such behavior of Ruppeiner curvature has also been reported for Takahashi gas \cite{rupp90}, finite one- and two-dimensional Ising models \cite{brody}, one-dimensional Potts model \cite{potts}, and two-dimensional ideal anyon gas \cite{anyon}. Whether there are deeper resemblances between these models and charged Ho\v{r}ava-Lifshitz black holes is left for future studies.

Now, we study the Legendre invariant metric proposed by Quevedo \cite{que06}. In this formalism, which is known as geometrothermodynamics, we consider a $(2n+1)$-dimensional thermodynamic phase space ${\mathcal T}$, with coordinates $\{\Phi, E^{a}, I^{a}\}$, $a=1, \ldots, n$. $\Phi$ represents thermodynamic potential, and $E^{a}$ and $I^{a}$ are extensive and intensive variables, respectively. The positive integer $n$ indicates the number of degrees of freedom of thermodynamic system under study. We also consider the thermodynamic equilibrium state subspace of  ${\mathcal T}$, which is denoted by  ${\mathcal E}$ and defined by the embedding map
\begin{equation}
\varphi: \{E^{a}\}\mapsto \{\Phi, E^{a}, I^{a}\},
\end{equation}
with $\Phi=\Phi(E^{a})$. Furthermore, we introduce the fundamental Gibbs 1-form on ${\mathcal T}$ as
\begin{equation}
\Theta=d\Phi -\delta_{ab}E^{a}I^{b}, \qquad \delta_{ab}= diag(1, 1, \ldots, 1),
\end{equation}
whose projection on ${\mathcal E}$ vanishes, giving the first law and conditions for thermodynamic equilibrium
\begin{equation}
d\Phi=\delta_{ab}E^{a}I^{b}, \quad \frac{\partial \Phi}{\partial E^{a}}=\delta_{ab}I^{b}.
\end{equation}

In \cite{que11} Quevedo \textit{et al.} have proposed the Legendre invariant metric
\begin{eqnarray}
G&=&\Theta^{2}+(\delta_{ab}E^{a}I^{b})(\eta_{cd}dE^{c}dI^{d}), \\
&&\eta_{ab}= diag(-1, 1, \ldots, 1), \nonumber
\end{eqnarray}
for the manifold ${\mathcal T}$, which induces the metric
\begin{equation}
g=(E^{c}\frac{\partial \Phi}{\partial E^{c}})(\eta_{ab}\delta^{bc}\frac{\partial^{2} \Phi}{\partial E^{c}\partial E^{d}}dE^{a}dE^{d})
\label{inducedmet}
\end{equation}
on the submanifold  ${\mathcal E}$, and can be used to investigate systems undergoing second order phase transition.

For charged black holes in extended phase space in an ensemble that both charge and pressure could vary, we take $\bar{M}=M-\Phi Q-PV$ as the thermodynamic potential. By using the first law we have $d\bar{M}=TdS-Qd\Phi-PdV$. So it would be convenient to take $E^{c}=\{S, \Phi, V\}$, and the Quevedo metric is found to be 

\begin{equation}
g^{Q}=(ST-Q\Phi-PV) \left( \begin{array}{ccc}
-\frac{\partial^{2} \bar{M}}{\partial S^{2}} & 0 & 0 \\
0 & \frac{\partial^{2} \bar{M}}{\partial \Phi^{2}} & \frac{\partial^{2} \bar{M}}{\partial V \partial \Phi} \\
0 & \frac{\partial^{2} \bar{M}}{\partial V \partial \Phi} & \frac{\partial^{2} \bar{M}}{\partial V^{2}} \end{array} \right).
\end{equation}

\begin{figure}[t]
	\centering
	\includegraphics[scale=.9]{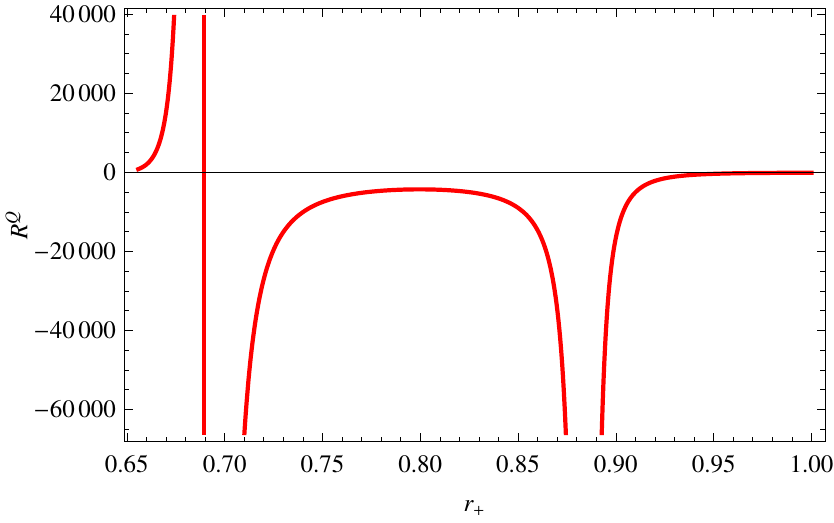}
	\caption[]{Scalar curvature of Quevedo metric as a function of horizon radius $r_{+}$ for $k=-1$, $\Phi=1$, and $V=-1$.}
	\label{qricci}
\end{figure}

We have plotted the scalar curvature of Quevedo metric in Fig. \ref{qricci}. Like the case of Weinhold and Ruppeiner curvature, the scalar curvature of Quevedo metric diverges at the point of second order phase transition ($r_{+}=0.88443$). But Quevedo curvature has another divergent point at $r_{+}=0.68953$ for which the specific heat $C_{\Phi,V}$ is finite. This result shows that the scalar curvature of Quevedo metric can not be used to uniquely predict the second order phase transition of extended charged Ho\v{r}ava-Lifshitz black hole. A similar result has been obtained for phantom Reissner-Nordestr\"{o}m-AdS black hole \cite{mansoori}.

\section{Conclusions} \label{seccon}

By treating cosmological constant as thermodynamic pressure and taking its conjugate quantity as thermodynamic volume, black holes become more like laboratory thermodynamic systems. In this paper, we have studied charged Ho\v{r}ava-Lifshitz black hole in three different ensembles; ensembles with fixed pressure and fixed/varying charge and an ensemble with both pressure and charge taken to be fixed. We have found that for black holes with hyperbolic event horizon ($k=-1$), the specific heat has a divergent point in all of these ensembles.

On the other hand, it is known that, in general relativity, the specific heat of charged black hole has a divergent point only for the case of spherical horizon ($k=1$) \cite{kubizpv}. In fact, there exist a duality between the solutions of (charged) Ho\v{r}ava-Lifshitz gravity  with  $k=-1, 0$, and $1$ and that of Einstein(-Maxwell) theory  with  $k=1, 0$, and $-1$, respectively. We have shown that the reason of this duality is the presence of higher derivative terms in the Lagrangian of Ho\v{r}ava-Lifshitz theory. More on the similarities between the charged black holes in Ho\v{r}ava-Lifshitz theory and Einstein gravity have been investigated in \cite{majhicqg}. By calculating the critical exponents of charged Ho\v{r}ava-Lifshitz black hole it has been shown that the charged black hole in Ho\v{r}ava-Lifshitz gravity with hyperbolic horizon is in the same universality class as that in Einstein gravity with horizon of spherical topology.

Ehrenfest equations in extended phase space are presented in the three different ensembles. We showed that, given the continuity of the first derivatives of the Gibbs free energy, all of the extended Ehrenfest equations are satisfied even at the point where the specific heat diverges. The importance of (extended) Ehrenfest equations in black hole thermodynamics is that they express the variation of pressure and charge in terms of other thermodynamic quantities.

It is known in the context of non-equilibrium thermodynamics, that vitrification proceeds in some finite temperature interval \cite{jackle}, and it has been shown in \cite{schmelzer}, that this results in a value for PD ratio larger than one for glassy phase transitions. In this paper, we have shown that one obtains the value of PD ratio equal to one, directly from the fact that the specific heat diverges at a sharp temperature. So, the divergence of the specific heat at a sharp temperature in black hole thermodynamics rules out the possibility of glassy phase transition.

We have analyzed local stability of charged extended Ho\v{r}ava-Lifshitz black hole. If the black hole is locally unstable, any perturbations in thermodynamic parameters will drop the black hole out of its equilibrium state. By requiring the concavity of entropy function or equivalently the convexity of the mass of the black hole, we have found the conditions under which the black hole is locally thermodynamically stable. We have shown that the stable charged extended Ho\v{r}ava-Lifshitz black hole lies in a region of the phase space that the thermodynamic volume is negative.

We have also studied the phase transition of charged extended Ho\v{r}ava-Lifshitz black hole in thermodynamic geometry. We have examined Weinhold, Ruppeiner, and Quevedo metric and we have shown that the scalar curvature in all of these metrics have a divergent point at the point of second order phase transition. Ruppeiner curvature diverges for the extremal black hole as well. It also passes through $R^{R}=0$ and changes its sign, a behavior which is also reported for two-dimensional ideal anyon gas \cite{anyon} and some others thermodynamic systems \cite{rupp90,brody,potts}. The problem arises for the Quevedo metric which has another divergent point at the point where no phase transition occurs. It is the matter of future studies to investigate other black holes in extended phase space in the context of geometrothermodynamics to see whether they produce similar results for phase transitions.

\section*{Appendix: Extended Ehrenfest equations}
\renewcommand{\theequation}{A.\arabic{equation}}

Here, we extend Ehrenfest equations to obtain nine relations between thermodynamic quantities which are true at the point in which the specific heat diverges. Our approach is based on the recently developed method of \cite{mine}. We consider the general case of charged extended black holes in an ensemble with fixed charge and pressure. In this ensemble the Gibbs free energy is defined as $G=H-TS=M-TS$. So, by using the first law (\ref{1stlaw}), we have
\begin{equation}
dG=-SdT+VdP+\Phi dQ. \label{dgibbs}
\end{equation}
So
\begin{equation}
S=-(\frac{\partial G}{\partial T})_{P,Q}, \qquad V=(\frac{\partial G}{\partial P})_{T,Q}, \qquad \Phi=(\frac{\partial G}{\partial Q})_{T,P}. \nonumber
\end{equation}
Maxwell relations are obtained straightforwardly
\begin{eqnarray}
(\frac{\partial S}{\partial P})_{T,Q}&=&-(\frac{\partial V}{\partial T})_{P,Q}, \nonumber\\
(\frac{\partial S}{\partial Q})_{T,P}&=&-(\frac{\partial \Phi}{\partial T})_{P,Q}, \\
(\frac{\partial \Phi}{\partial P})_{T,Q}&=&(\frac{\partial V}{\partial Q})_{T,P}. \nonumber
\end{eqnarray}

The first derivatives of the Gibbs free energy are continuous at the divergent point of the specific heat. So, $S_{1}=S_{2}$, $V_{1}=V_{2}$, and $\Phi_{1}=\Phi_{2}$, where the indices $1$ and $2$ denote the state before and after the transition. Then, one can conclude that
\begin{equation}
dS_{1}=dS_{2}, \qquad dV_{1}=dV_{2}, \qquad d\Phi_{1}=d\Phi_{2}.
\end{equation}

By expressing entropy $S$ as a function of temperature $T$, pressure $P$, and charge $Q$, we would have
\begin{equation}
dS=\frac{C_{P,Q}}{T}dT-V\alpha dP-\Phi\alpha^{'}dQ, \label{ds}
\end{equation}
where we have used the definition of the specific heat $C_{P,Q}=T(\frac{\partial S}{\partial T})_{P,Q}$, volume expansion coefficient $\alpha=-\frac{1}{V}(\frac{\partial S}{\partial P})_{T,Q}=\frac{1}{V}(\frac{\partial V}{\partial T})_{P,Q}$, and $\alpha^{'}=-\frac{1}{\Phi}(\frac{\partial S}{\partial Q})_{P,T}=\frac{1}{\Phi}(\frac{\partial \Phi}{\partial T})_{P,Q}$. Since temperature $T$, volume $V$, and electric potential $\Phi$ are continuous and $dS_{1}=dS_{2}$, Eq. (\ref{ds}) gives
\begin{eqnarray}
V(\alpha_{2}-\alpha_{1})(\frac{dP}{dT})_{S}+\Phi(\alpha^{'}_{2}-\alpha^{'}_{1})(\frac{dQ}{dT})_{S} \nonumber\\
-\frac{(C_{P,Q})_{2}-(C_{P,Q})_{1}}{T}=0. \label{2fold1}
\end{eqnarray}
The above equation can be rearranged to yield
\begin{equation}
(\frac{dP}{dT})_{S}=\frac{(C_{P,Q})_{2}-(C_{P,Q})_{1}}{TV(\alpha_{2}-\alpha_{1})} -\frac{\Phi(\alpha^{'}_{2}-\alpha^{'}_{1})}{V(\alpha_{2}-\alpha_{1})}(\frac{dQ}{dT})_{S}. \label{dpdts}
\end{equation}
Pressure $P$ can be expressed as a function of $T$, $S$, and $Q$ at least in principle, so
\begin{equation}
dP=(\frac{\partial P}{\partial T})_{S,Q}dT+(\frac{\partial P}{\partial S})_{T,Q}dS +(\frac{\partial P}{\partial Q})_{T,S}dQ.
\end{equation}
For $dS=0$ we would obtain
\begin{equation}
(\frac{dP}{dT})_{S}=(\frac{\partial P}{\partial T})_{S,Q}+(\frac{\partial P}{\partial Q})_{T,S}(\frac{dQ}{dT})_{S}.
\end{equation}
Comparing this equation with (\ref{dpdts}) we find the first and second extended Ehrenfest equations
\begin{eqnarray}
(\frac{\partial P}{\partial T})_{S,Q}&=&\frac{(C_{P,Q})_{2}-(C_{P,Q})_{1}}{TV(\alpha_{2}-\alpha_{1})}, \label{1st} \\
(\frac{\partial P}{\partial Q})_{T,S}&=&-\frac{\Phi(\alpha^{'}_{2}-\alpha^{'}_{1})}{V(\alpha_{2}-\alpha_{1})}. \label{2nd}
\end{eqnarray}

Equation (\ref{2fold1}) can also be rearranged to give
\begin{equation}
(\frac{dQ}{dT})_{S}=\frac{(C_{P,Q})_{2}-(C_{P,Q})_{1}}{T\Phi(\alpha^{'}_{2}-\alpha^{'}_{1})}-\frac{V(\alpha_{2}-\alpha_{1})}{\Phi(\alpha^{'}_{2}-\alpha^{'}_{1})}(\frac{dP}{dT})_{S}. \label{dqdts}
\end{equation}
For $Q$ as a function of $T$, $S$, and $P$, one can write, for constant entropy,
\begin{equation}
(\frac{dQ}{dT})_{S}=(\frac{\partial Q}{\partial T})_{S,P}+(\frac{\partial Q}{\partial P})_{T,S}(\frac{dP}{dT})_{S}.
\end{equation}
By comparing this equation with (\ref{dqdts}) we would find the third and fourth extended Ehrenfest equations
\begin{eqnarray}
(\frac{\partial Q}{\partial T})_{S,P}&=&\frac{(C_{P,Q})_{2}-(C_{P,Q})_{1}}{T\Phi(\alpha^{'}_{2}-\alpha^{'}_{1})}, \label{3rd} \\
(\frac{\partial Q}{\partial P})_{T,S}&=&-\frac{V(\alpha_{2}-\alpha_{1})}{\Phi(\alpha^{'}_{2}-\alpha^{'}_{1})}. \label{4th}
\end{eqnarray}

For $V$ as a function of $T$, $P$, and $Q$, we have
\begin{equation}
dV=V\alpha dT+V\kappa dP+\Phi\kappa^{'}dQ,
\end{equation}
where $\kappa=-\frac{1}{V}(\frac{\partial V}{\partial P})_{T,Q}$ is the isothermal compressibility and $\kappa^{'}=\frac{1}{\Phi}(\frac{\partial V}{\partial Q})_{T,P}=\frac{1}{\Phi}(\frac{\partial \Phi}{\partial P})_{T,Q}$. Following the same procedure as above and $dV_{1}=dV_{2}$, we find four other extended Ehrenfest equations
\begin{eqnarray}
(\frac{\partial P}{\partial T})_{V,Q}&=&\frac{\alpha_{2}-\alpha_{1}}{\kappa_{2}-\kappa_{1}}, \label{5th} \\
(\frac{\partial P}{\partial Q})_{T,V}&=&\frac{\Phi(\kappa^{'}_{2}-\kappa^{'}_{1})}{V(\kappa_{2}-\kappa_{1})}, \label{6th} \\
(\frac{\partial Q}{\partial T})_{V,P}&=&-\frac{V(\alpha_{2}-\alpha_{1})}{\Phi(\kappa^{'}_{2}-\kappa^{'}_{1})}, \label{7th} \\
(\frac{\partial Q}{\partial P})_{T,V}&=&\frac{V(\kappa_{2}-\kappa_{1})}{\Phi(\kappa^{'}_{2}-\kappa^{'}_{1})}. \label{8th}
\end{eqnarray}

A general expression for $\Phi$ as a function of $T$, $Q$, and $P$ would give
\begin{equation}
d\Phi=\Phi\alpha^{'}dT+\Phi\kappa^{'}dP+\Phi\chi dQ,
\end{equation}
in which we have used $\chi=\frac{1}{\Phi}(\frac{\partial\Phi}{\partial Q})_{T,P}$. By the same calculations we find the following equations
\begin{eqnarray}
(\frac{\partial P}{\partial T})_{\Phi,Q}&=&-\frac{\alpha^{'}_{2}-\alpha^{'}_{1}}{\kappa^{'}_{2}-\kappa^{'}_{1}}, \label{9th} \\
(\frac{\partial P}{\partial Q})_{T,\Phi}&=&-\frac{\chi_{2}-\chi_{1}}{\kappa^{'}_{2}-\kappa^{'}_{1}}, \label{10th} \\
(\frac{\partial Q}{\partial T})_{\Phi,P}&=&-\frac{\alpha^{'}_{2}-\alpha^{'}_{1}}{\chi_{2}-\chi_{1}}, \label{11th} \\
(\frac{\partial Q}{\partial P})_{T,\Phi}&=&-\frac{\kappa^{'}_{2}-\kappa^{'}_{1}}{\chi_{2}-\chi_{1}}. \label{12th}
\end{eqnarray}

Not all of the twelve equations we have obtained, are independent. In fact Eqs. (\ref{4th}), (\ref{8th}), and (\ref{12th}) are respectively the reverse of (\ref{2nd}), (\ref{6th}), and (\ref{10th}). Therefore, a total of nine independent extended Ehrenfest equations in canonical ensemble remain.


\begin{thebibliography}{}
	
	\bibitem{bhexp}
	S.W. Hawking, Nature {\bf 248}, 30 (1974)
	\bibitem{bhent}
	J.D. Bekenstein, Phys. Rev. D {\bf 7}, 2333 (1973)
	\bibitem{hptr}
	S.W. Hawking, D.N. Page, Commun. Math. Phys. {\bf 87}, 577 (1983)
	\bibitem{adscft}
	E. Witten, Adv. Theor. Math. Phys. {\bf 2}, 505 (1998)
	\bibitem{lis11}
	T.K. Dey, S. Mukherji, S. Mukhopadhyay, S. Sarkar, JHEP {\bf 04}, 014 (2007)
	\bibitem{lis12}
	Y.S. Myung, Y.W. Kim, Y.J. Park, Phys. Rev. D {\bf 78}, 084002 (2008)
	\bibitem{lis13}
	D. Anninos, G. Pastras, JHEP {\bf 07}, 030 (2009)
	\bibitem{lis14}
	S. Carlip, S. Vaidya, Class. Quant. Grav. {\bf 20}, 3827 (2003)
	\bibitem{lis15}
	B.N.M. Carter, I.P. Neupane, Phys. Rev. D {\bf 72}, 043534 (2005)
	\bibitem{lis16}
	S. Fernando, Phys. Rev. D {\bf 74}, 104032 (2006)
	\bibitem{lis17}
	A. Chamblin, R. Emparan, C.V. Johnson, R.C. Myers, Phys. Rev. D {\bf 60}, 064018 (1999)
	\bibitem{lis18}
	A. Chamblin, R. Emparan, C.V. Johnson, R.C. Myers, Phys. Rev. D {\bf 60}, 104026 (1999)
	\bibitem{lis19}
	M.M. Caldarelli, G. Cognola, D. Klemm, Class. Quant. Grav. {\bf 17}, 399 (2000)
	\bibitem{lis110}
	R. Banerjee, S.K. Modak, D. Roychowdhury, JHEP {\bf 10}, 125 (2012)
	\bibitem{traschen}
	D. Kastor, S. Ray, J. Traschen, Class. Quant. Grav. {\bf 26}, 195011 (2009)
	\bibitem{dolan5023}
	B.P. Dolan, Class. Quant. Grav. {\bf 28}, 235017 (2011)
	\bibitem{kubizpv}
	D. Kubiz\v{n}\'{a}k, R.B. Mann, JHEP {\bf 07}, 033 (2012)
	\bibitem{mann6251}
	S. Gunasekaran, D. Kubiz\v{n}\'{a}k, R.B. Mann, JHEP {\bf 11}, 110 (2012)
	\bibitem{mannsherkat1}
	N. Altamirano, D. Kubiz\v{n}\'{a}k, R.B. Mann, Z. Sherkatghanad, Class. Quant. Grav. {\bf 31}, 042001 (2014)
	\bibitem{moliuprd}
	J.X. Mo, W.B. Liu, Phys. Rev. D {\bf 89}, 084057 (2014)
	\bibitem{moepl}
	J.X. Mo, EPL (Europhysics Letters) {\bf 105}, 20003 (2014)
	\bibitem{moliuplb}
	J.X. Mo, W.B. Liu, Phys. Lett. B {\bf 727}, 336 (2013)
	\bibitem{horava1}
	P. Ho\v{r}ava, Phys. Rev. D {\bf 79}, 084008 (2009)
	\bibitem{horava2}
	P. Ho\v{r}ava, JHEP {\bf 03}, 020 (2009)
	\bibitem{lmp}
	H. Lu, J. Mei, C.N. Pope, Phys. Rev. Lett. {\bf 103}, 091301 (2009)
	\bibitem{ccoprd}
	R.G. Cai, L.M. Cao, N. Ohta, Phys. Rev. D {\bf 80}, 024003 (2009)
	\bibitem{ks}
	A. Kehagias, K. Sfetsos, Phys. Lett. B {\bf 678}, 123 (2009)
	\bibitem{ghodsi}
	A. Ghodsi, E. Hatefi, Phys. Rev. D {\bf 81}, 044016 (2010)
	\bibitem{myung}
	Y.S. Myung, Y.W. Kim, Eur. Phys. J. C {\bf 68}, 265 (2010)
	\bibitem{castillo}
	A. Castillo, A. Larranaga, Electron. J. Theor. Phys. {\bf 8} 83 (2011)
	\bibitem{majhiplb}
	B.R. Majhi, Phys. Lett. B {\bf 686}, 49 (2010)
	\bibitem{mojhep}
	J.X. Mo, X.X. Zeng, G.Q. Li, X. Jiang, W.B. Liu, JHEP {\bf 10}, 056 (2013)
	\bibitem{suresh}
	J. Suresh, R. Tharanath, V.C. Kuriakose, JHEP {\bf 01}, 019 (2015)
	\bibitem{cao5044}
	Q.J. Cao, Y.X. Chen, K.N. Shao, Phys. Rev. D {\bf 83}, 064015 (2011)
	\bibitem{majhicqg}
	B.R. Majhi, D. Roychowdhury, Class. Quantum Grav. {\bf 29}, 245012 (2012)
	\bibitem{sadeghi}
	J. Sadeghi, K. Jafarzade, B. Pourhassan, Int. J. Thoer. Phys {\bf 51}, 3891 (2012)
	\bibitem{mine}
	M.B. Jahani Poshteh, B. Mirza, F. Oboudiat, Int. J. Mod. Phys. D {\bf 24}, 1550029 (2015)
	\bibitem{weinhold}
	F. Weinhold, J. Chem. Phys. {\bf 63}, 2479 (1975)
	\bibitem{rupp79}
	G. Ruppeiner, Phys. Rev. A {\bf 20}, 1608 (1979)
	\bibitem{que06}
	H. Quevedo, J. Math. Phys. {\bf 48}, 013506 (2007)
	\bibitem{nopv}
	J.X. Mo, Astrophys. Space Sci. {\bf 356}, 319 (2015)
	\bibitem{brill}
	D.R. Brill, J. Louko, P. Peldan, Phys. Rev. D {\bf 56}, 3600 (1997)
	\bibitem{pd}
	I. Prigogine, R. Defay, {\it Chemical Thermodynamics} (Longmans, London, 1954)
	\bibitem{jackle}
	J. J\"{a}ckle, Rep. Prog. Phys. {\bf 49}, 171 (1986)
	\bibitem{nie}
	T.M. Nieuwenhuizen, Phys. Rev. Lett. {\bf 79}, 1317 (1997)
	\bibitem{monteiro}
	R.J.F. Monteiro, Classical and thermodynamic stability of black holes arXiv:1006.5358[hep-th]
	\bibitem{gabriel}
	S. Gilbert, "Linear Algebra and its Applications" Brooks/Cole Thomson Learning, Inc. (1988).
	\bibitem{salamon}
	P. Salamon, J.D. Nulton, E. Ihrig, J. Chem. Phys. {\bf 80}, 436 (1984)
	\bibitem{rupp90}
	G. Ruppeiner, J. Chance, J. Chem. Phys. {\bf 92}, 3700 (1990)
	\bibitem{brody}
	D.C. Brody, A. Ritz, J. Geom. Phys. {\bf 47}, 207 (2003)
	\bibitem{potts}
	B.P. Dolan, D.A. Johnston, R. Kenna, J. Phys. A {\bf 35}, 9025 (2002)
	\bibitem{anyon}
	B. Mirza, H. Mohammadzadeh, Phys. Rev. E {\bf 80}, 011132 (2009)
	\bibitem{que11}
	H. Quevedo, A. S\'{a}nchez, S. Taj, A. V\'{a}zquez, Gen. Relativ. Gravit. {\bf 43}, 1153 (2011)
	\bibitem{mansoori}
	S.A.H Mansoori, B. Mirza, Eur. Phys. J. C {\bf 74}, 1 (2014)
	\bibitem{schmelzer}
	J.W. Schmelzer, I. Gutzow, J. Chem. Phys. {\bf 70}, 184511 (2006)

\end{thebibliography}
\end{document}